# On some common compressive sensing recovery algorithms and applications
## - Review paper-

Andjela Draganić, Irena Orović, Srdjan Stanković

University of Montenegro, Faculty of Electrical Engineering, Podgorica, Montenegro

*Abstract: Compressive Sensing, as an emerging technique in signal processing is reviewed in this paper together with its' common applications. As an alternative to the traditional signal sampling, Compressive Sensing allows a new acquisition strategy with significantly reduced number of samples needed for accurate signal reconstruction. The basic ideas and motivation behind this approach are provided in the theoretical part of the paper. The commonly used algorithms for missing data reconstruction are presented. The Compressive Sensing applications have gained significant attention leading to an intensive growth of signal processing possibilities. Hence, some of the existing practical applications assuming different types of signals in real-world scenarios are described and analyzed as well.*

### 1. THE BASIC COMPRESSIVE SENSING CONCEPTS

Continuous time, bandlimited signals, sampled according to the Shannon-Nyquist sampling theorem, may produce a large number of samples to be further processed. Having in mind that the signal samples are acquired at the rate being at least twice the maximal signal frequency, the conventional sampling may be inefficient, especially in applications dealing with the high-frequency signals. Also, a large number of sensors required for acquisition may lead to large power consumption. Hence, the compression arises as a necessary step in conventional signal processing. Most of the signals we are dealing with contain redundant information, and this fact is exploited during the compression step. The compression discards certain percent of the samples in the sparse transformation domain, assuming that majority of samples are insignificant for signal analysis. The fact that the most signals exhibit sparsity in certain transformation domain is used in the Compressive Sensing (CS) theory [1]-[12]. Namely, one of the ideas behind the CS was to avoid compression after acquisition and to directly acquire data in the compressed form. In other words, the CS offers the possibility to acquire less data then it is commonly done, but still to be able to reconstruct the entire information afterwards. The missing signal information can also appear as a consequence of omitting samples that are exposed to different kinds of noise or losing some parts of the signal during the transmission. These missing samples can be recovered using the CS reconstruction algorithms [7]-[27].

Some of the concepts that are nowadays used within the CS approaches dates from the early seventies. The least square solutions, based on the norm minimization, are used by Claerbout and Muir in 1973 [11]. In 1986, Santosa and Symes proposed an application of the $\ell_1$-norm in recovering sparse spike trains. The $\ell_1$-minimization of the image gradient - total variation minimization, is proposed in 1990s by Rudin, Osher and Fatemi [17], for removing noise from images. In the early 2000s Blu, Marziliano, and Vetterli showed that the *K*-sparse signals can be sampled and recovered by using only 2*K* parameters. The idea of the CS starts to grow from the moment when it was shown that a small set of non-adaptive measurements can provide exact signal reconstruction, which proved the basic idea behind the data acquisition in the compressed form [3],[4]. Later, in [19], the CS is analyzed in terms of signal recovery when the missing samples are result of signal degradation due to the noise presence. The influence of the number of missing samples on the spectral signal representation is examined, and the reconstruction procedure is proposed.

Our focus in this paper is on the practical applications of the CS approaches. However, there are some specific requirements that are imposed to the measurements, in order to be able to apply CS signal reconstruction. Signal sparsity is one of the conditions required in CS approach, and can be satisfied in different domains: time, frequency or time-frequency domains [28]-[48]. This condition is valid for variety of real-world signals. The other condition is incoherence which will be explained later in the text.

There is a large number of CS applications, from those assuming one-dimensional signals to various image processing and video applications. Some of the CS applications are adopted to work in the real-time. The constant growth and development in the field of the CS applications aims to reduce the complexity of devices, to speed up the acquisition and transmission procedure and to decrease the power consumption. Since CS is used to extract as much as possible information from minimal available data, it is important to highlight the use in biomedicine, especially in Magnetic Resonance Imaging – MRI. By lowering the number of coefficients required for MR image reconstruction, the time of patient exposure to the MR device is reduced and consequently, the negative influence of the MR device is lower. Another useful application is in radar imaging, where the CS exploits the sparsity in the frequency domain. Furthermore, it is used in communication and network systems, sparse channel estimation, wireless sensor networks (WSNs), in cognitive radios for spectrum sensing, etc. Some of the applications will be described later in the text.

## 2. THE MATHEMATICAL BACKGROUND OF THE COMPRESSIVE SENSING CONCEPT

Assume that we are dealing with the signal **x** of length *N*, that is sparse in the transform domain (defined by the direct transformation $\Im$). Then, the vector of acquired samples **y** can be defined as [1]-[12]:

$$\mathbf{y} = \Omega \Im^{-1} \mathbf{X}, \qquad (1)$$

where matrix $\mathbf{\Omega}$ is used to randomly under-sample the observed signal. The signal sparsity is *K*, meaning that only *K* out of *N* coefficients from the transform domain are non-zero and we assume that only *M* out of *N* samples are acquired in the vector **y** (*M*<<*N* and *M*>2*K*). The vector **X** is the vector of the transform domain coefficients, i.e.:

$$\mathbf{X} = \Im \mathbf{x}, \qquad (2)$$

and $\Im^{-1}$ is the inverse transform matrix. Different transform domains can be used: discrete Fourier transform domain – DFT [3],[6],[12], discrete cosine transform domain – DCT [3],[6],[12],[46], wavelet domain, Hermite transform domain [48]-[55], time-frequency domain [56], etc. Apart from the sparsity, another important property is incoherence, which enables successful signal reconstruction from small set of acquired samples. Namely, the measurement matrix $\boldsymbol{\Omega}$ should be incoherent with the transform domain matrix $\Im$. The coherence between the two matrices represents the highest correlation between any two column/row vectors of the matrices. A measure of correlation between the two matrices is defined as follows [11],[12]:

$$\mu(\Omega, \Im) = \sqrt{N} \max_{k \geq 1, j \leq N} \left| \langle \Omega_k, \Im_j \rangle \right|, \quad (3)$$

where $N$ is a signal length, $\boldsymbol{\Omega}_k$ and $\Im_j$ are row and column vectors of the matrices $\boldsymbol{\Omega}$ and $\Im$, respectively. The coherence takes values from the interval:

$$1 \leq \mu(\Omega, \Im) \leq \sqrt{N}. \quad (4)$$

The value of the coherence is greater if the two matrices are more correlated. In the CS scenario the value of the coherence should be as low as possible.

The system of equations (1) can be written as follows:

$$\mathbf{y}_{M \times 1} = \mathbf{A}_{M \times N} \mathbf{X}_{N \times 1}, \quad (5)$$

where $\mathbf{A}$ denotes CS matrix: $\mathbf{A} = \boldsymbol{\Omega}\Im^{-1}$. The system is under-determined since it has $M$ equations and $N$ unknowns. Therefore, the optimization techniques are used in order to find an optimal solution for this system. Optimal solution is related to the signal sparsity – the sparsest solution is the optimal one.

There is a number of algorithms used to obtain a sparse solution of the system. Some of them are based on the convex optimization [1]-[7],[11],[12]: basis pursuit, Dantzig selector, and gradient-based algorithms. They provide high reconstruction accuracy, but they are computationally demanding. The commonly used and less computationally demanding are greedy algorithms – Matching Pursuit and Orthogonal Matching Pursuit [1]-[6],[8],[12]. Also, recently proposed threshold based algorithms provide high reconstruction accuracy with low computational complexity: (e.g. Iterative Hard Thresholding - IHT, Iterative Soft Thresholding - IST) [1],[6],[13], Automated Threshold Based Iterative Solution [12],[57], Adaptive Gradient-Based Algorithm, [3],[12],[26],[27], etc.

## 3. COMPRESSIVE SENSING ALGORITHMS

The sparsity of the signal can be defined as a number of nonzero elements within a vector. It can be described by using the $\ell_0$-norm [13]:

$$\|\mathbf{x}\|_0 = \lim_{p \to 0} \sum_{i=1}^{N} |x_i|^p = \sum_{i=1, x_i \neq 0}^{N} 1 = K, \quad (6)$$

and represents the cardinality of the support of $\mathbf{x}$:

$$\|\mathbf{x}\|_0 = \text{card}\{\text{supp}(\mathbf{x})\} \leq K. \qquad (7)$$

Therefore, the solution of the undetermined system of equations (5), in the cases when the signal **x** is sparse in the transform domain, can be reduced to the minimization of the $\ell_0$-norm, i.e.:

$$\min \|\mathbf{X}\|_0 \text{ subject to } \mathbf{y} = \mathbf{AX}. \qquad (8)$$

The $\ell_0$-norm is not feasible in practice, since small noise in the signal will be assumed as a non-zero sample. Therefore, the $\ell_1$-norm is commonly used. The optimization problem based on the $\ell_1$-norm is recast as follows [12],[13]:

$$\min \|\mathbf{X}\|_1 \text{ subject to } \mathbf{y} = \mathbf{AX}. \qquad (9)$$

In the sequel, some of the commonly used algorithms for sparse reconstruction are described.

### 3.1. Convex optimizations
#### Basis Pursuit and Basis Pursuit Denoising

The equation (8) represents a non-convex combinatorial optimization problem. Solution of this problem requires exhaustive searches over subsets of columns of the matrix **A**. For a *K*-sparse signal of length *N*, the total number of *K*-position subsets is $\binom{N}{K}$, which is not computationally feasible. Other approach solves a convex optimization problem through a linear programming, which is computationally more efficient. Commonly used convex optimization algorithms are: Basis Pursuit, Basis Pursuit De-Noising (BPDN), Least Absolute Shrinkage and Selection Operator (LASSO), Least Angle Regression (LARS), etc.

The approach based on the convex $\ell_1$-minimization that provides near optimal solution, can be defined as:

$$\min \|\mathbf{X}\|_1 \text{ subject to } \mathbf{y} = \mathbf{AX}. \qquad (10)$$

This approach is known as a Basis Pursuit (BP) [6],[12]. It aims in decomposition of a signal into a superposition of dictionary elements that have the smallest $\ell_1$-norm of the coefficients. BP can be solved by using a primal-dual interior point method. The problem (10) can be recast as follows, in the case of real **y**, **A** and **X** [12]:

$$\min_t \sum t, \text{ subject to } -t \leq \mathbf{X} \leq t, \mathbf{y} = \mathbf{AX}, \qquad (11)$$

where variable *t* is introduced to avoid absolute value in $\|\mathbf{X}\|_1 = \sum_{i=1}^{N}|\mathbf{X}_i|$. The steps of the primal-dual interior point method are described within the Algorithm 1. In the cases of noisy measurements, **y=AX+n**, where **n** denotes noise and $\|\mathbf{n}\|_2 \leq \varepsilon$, the optimization problem is known as Basis Pusuit Denoising (BPD) and is defined as [6]:

$$\min \|\mathbf{X}\|_1 \text{ subject to } \|\mathbf{y} - \mathbf{AX}\|_2 \leq \varepsilon. \qquad (12)$$

Step directions for the Algorithm 1 are obtained by finding the first derivatives of $\Lambda$ in terms of its arguments. Step lengths are calculated using the backtracking line search [12]. For example, a new value for **X** is obtained as **X=X**+*u*($\Delta$**X**).

**Algorithm 1: Primal-dual interior point method**

- Set $\mathbf{X} = \mathbf{X_0} = \mathbf{A}^T \mathbf{y}$, for the known measurement vector $\mathbf{y}$.
- Set $t_0 = \gamma |\mathbf{X_0}| + \lambda \max\{|\mathbf{X_0}|\}$. Parameters $\gamma$ and $\lambda$ are user-defined.
- The next step is forming a Lagrangian function:

$$\Lambda\left(\mathbf{X}, t, g, -\frac{1}{\mathbf{X_0}-t_0}, \frac{1}{\mathbf{X_0}+t_0}\right) = f(t) + g(\mathbf{AX}-\mathbf{y}) - \frac{\mathbf{X}-t}{\mathbf{X_0}-t_0} - \frac{\mathbf{X}+t}{\mathbf{X_0}+t_0}, \text{ where } g = -\mathbf{A}\left(\frac{-1}{\mathbf{X_0}-t_0} + \frac{1}{-\mathbf{X_0}-t_0}\right).$$

- Update each argument of the Lagrangian function by step direction ($\Delta$) and step length ($u$).

*Adaptive gradient based algorithm*

Adaptive gradient based algorithm proposed in [26], belongs to the group of convex optimization approaches. It starts from the chosen initial values of the available signal samples. The initial value is iteratively changed for $+\Delta$ and $-\Delta$, and the concentration improving is measured in the sparsity domain. The gradient vector, used to update the signal values, is obtained as a difference between the $\ell_1$-norms of the vectors changed for $+\Delta$ and changed for $-\Delta$. This gradient value is used to update the values of the missing samples.

The performance of this algorithm can be efficient even for the signals that are not strictly sparse. The algorithm for both 1D and 2D cases is summarized in the Algorithm 2.

**Algorithm 2: Adaptive gradient based algorithm**

**Input**: set of the positions of the available samples $\mathbf{\Omega_a}$ and set of the missing samples position: $\mathbf{\Omega_m} = \mathbf{N} \setminus \mathbf{\Omega_a}$; measurement vector $\mathbf{y}$; In the 1D signal case, $n=n$, while in the 2D signal case $n=(n_x, n_y)$

- Set $\mathbf{y}^{(0)}(n) \leftarrow \mathbf{y}(n)$ and $k \leftarrow 0$
- Set $\Delta \leftarrow \max|\mathbf{y}^{(0)}(n)|$
- **repeat**
- set $\mathbf{y}_p(n) \leftarrow \mathbf{y}^{(k)}(n)$
- **repeat**
- $k \leftarrow k+1$
- **for** $t \in \mathbf{N}$ **do**
- **if** $t \in \mathbf{\Omega}_m$ **then**

$X^+(f) \leftarrow \Im\{\mathbf{y}^{(k)}(n) + \Delta\}$,   ($f = f$ in the 1D case; $f = (f_1, f_2)$ in the 2D case);
$X^-(f) \leftarrow \Im\{\mathbf{y}^{(k)}(n) - \Delta\}$,   ($\Im$ DFT – 1D case; 2D DFT – 2D case)

$\mathbf{\Gamma}^{(k)}(t) \leftarrow \|X^+\|_1 - \|X^-\|_1$,

 **else**

$\mathbf{\Gamma}^{(k)}(t) \leftarrow 0$

- **end if**

$\mathbf{y}^{(k+1)}(t) \leftarrow \mathbf{y}^{(k)}(t) - \mathbf{\Gamma}^{(k)}(t)$

- **end for**

$$\beta_k = \arccos \frac{\langle \mathbf{\Gamma}^{k-1} \mathbf{\Gamma}^k \rangle}{\|\mathbf{\Gamma}^{k-1}\|_2^2 \|\mathbf{\Gamma}^k\|_2^2}$$

- **until** $\beta_k < 170°$
- $\Delta \leftarrow \Delta / \sqrt{10}$
- $R = 10 \log_{10} \left( \sum_{n \in \mathbf{\Omega}_m} |\mathbf{y}_p(n) - \mathbf{y}^{(k)}(n)|^2 / \sum_{n \in \mathbf{\Omega}_m} |\mathbf{y}^{(k)}(n)|^2 \right)$ **until** $R < R_{\max}$ – required precision
- **return** $\mathbf{y}^{(k)}(n)$

**Output**: reconstructed signal $\mathbf{y}^{(k)}(n)$

## 3.2. *Greedy algorithms*

The greedy algorithms represents the second group of algorithms used to obtain the sparsest solution of the system (5). These algorithms are less computationally complex and therefore much faster compared to the $\ell_1$-norm based optimization techniques, but are also less precise. The greedy algorithms are based on finding the elements of the transform matrix called dictionary that best matches the signal through iterations. Commonly used greedy algorithms are Matching Pursuit (MP), Orthogonal Matching Pursuit (OMP), Compressive Sampling Matching Pursuit (CoSaMP), etc.

The procedure for the OMP algorithm is described within the Algorithm 3.

## 3.3. *Threshold based algorithms*
### *Iterative hard and soft thresholding*

Thresholding algorithms are based on an adaptive threshold applied within several iterations. They are much faster than algorithms based on convex relaxation. An iteration can be described in terms of threshold function as [1],[6],[11],[13]:

$$x_i = T_\varepsilon \left( f(X_{i-1}) \right). \qquad (13)$$

The thresholding function is denoted as $T_\varepsilon$, while $f$ is the function that modifies the output of the previous iterate and **X** is a sparse vector. The signal can be recovered from its measurement by using hard or soft thresholding. Therefore, there are two types of iterative thresholding algorithms: iterative hard thresholding (IHT) and iterative soft thresholding (IST).

IHT algorithm sets all but the *K* largest components, in terms of signal **X** magnitudes, to zero. The hard thresholding function $H_K$ is defined as [6],[11],[13]:

$$H_K(\mathbf{X}) = \begin{cases} X_i, & |X_i| > \varepsilon \\ 0, & \text{otherwise} \end{cases}. \qquad (14)$$

The ε is the *K* largest component of **X** [6]. The algorithm is summarized within the Algorithm 4 [11]. Soft thresholding function is applied to each element of the vector **X** and is defined as [6]:

$$S_\lambda(X_i) = \begin{cases} X_i - \lambda, & \mathbf{X} > \lambda \\ 0, & |X_i| < \lambda \\ X_i + \lambda, & \mathbf{X} < -\lambda \end{cases}. \qquad (15)$$

**Algorithm 3: Orthogonal matching pursuit**
- **Input:** Compressive sensing matrix $\mathbf{A} = \mathbf{\Omega}\mathbf{\Im}$, measurement vector **y**
- Initialization of the variables:
  - initial residual $\mathbf{r}_0 = \mathbf{y}$; initial solution $\mathbf{X}_0 = 0$; matrix of chosen atoms $\mathbf{\Upsilon}_0 = [\,]$.
- Do following steps until the stopping criterion is met:
  - $\omega_n = \underset{i=1,\ldots,M}{\arg\max} \left| \langle \mathbf{r}_{n-1}, \mathbf{A}_i \rangle \right|$    - finding maximum correlation column
  - $\mathbf{\Upsilon}_n \leftarrow \left[ \mathbf{\Upsilon}_{n-1} \; \mathbf{A}_{\omega_n} \right]$    - update matrix of chosen atoms
  - $\mathbf{X}_n = \underset{\mathbf{X}}{\arg\min} \left\| \mathbf{r}_{n-1} - \mathbf{\Upsilon}_n \mathbf{X}_{n-1} \right\|_2^2$    - solving least square problem
  - $\mathbf{r}_n = \mathbf{r}_{n-1} - \mathbf{\Upsilon}_n \mathbf{X}_{n-1}$    - residual update
  - $n = n+1$
- **Output:** $\mathbf{X}_P$ and $\mathbf{r}_P$, where *P* denotes number of iterations.

**Algorithm 4: Iterative hard thresholding**

**Input:** signal sparsity K, transform matrix $\mathfrak{I}$, measurement matrix $\Omega$, CS matrix **A**, measurement vector **y**

**Output:** an approximation of the signal $\hat{\mathbf{X}}$

$\mathbf{X}_0 \leftarrow 0$

**for** $i=1,\ldots,$ until stopping criterion is met **do**

$\mathbf{X}_i \leftarrow H_K\left(\mathbf{X}_{i-1} + \mathbf{A}^T(\mathbf{y} - \mathbf{A}\mathbf{X}_{i-1})\right)$

**end for**

**return** $\hat{\mathbf{X}} \leftarrow \mathbf{X}_i$

*Automated threshold based solution*

A non-iterative and iterative threshold based solutions for sparse signal reconstruction are proposed in [98]. The proposed solutions are based on the model of noise appearing as a consequence of missing samples. By using a predefined probability of error *P*, a general threshold *T* that separates signal components from spectral noise in the transform domain is defined.

**Algorithm 5: Automated threshold based iterative solution**

**Input:** $M$, $N$, **y**, $\Omega_a = \{n_1,\ldots,n_M\}$, $\mathfrak{I}$, $\Phi$, $\mathbf{A} = \Phi\mathfrak{I}$, $\sigma_N$.

o  Set $\mathbf{k} = \varnothing$ ;

**for** $i=1$ : $i=i+1$: **until** all components are detected

  o Calculate variance: $\sigma^2 = M\dfrac{N-M}{N-1}\sum_{i=1}^{M}\dfrac{y(i)^2}{M}$ ;

  o For a given $P$ calculate: $T = \sqrt{-\sigma^2 \log\left(1 - P(T)^{1/N}\right)}$ ;

  o Calculate the initial DFT vector $\mathbf{X}_i$: $\mathbf{X}_i = \mathbf{y}(\Phi\mathfrak{I}^{-1})$ ;

  o Update set **k**: $\mathbf{k} = \mathbf{k} \cup \arg\{|\mathbf{X}_i| > T/N\}$ ;

  o Calculate $\mathbf{F} = \left(\mathbf{A}^H \mathbf{A}\right)^{-1} \mathbf{A}\mathbf{y}$ ;         (CS matrix **A** contains rows defined by the set **k**, and *M* columns of the DFT matrix)

  o Update **y**: $\forall k \in \mathbf{k}$: $\mathbf{y} = \mathbf{x}(\Omega_a) - X(k)e^{j\frac{2\pi k \Omega_a}{N}}$ ;

  o Update the initial DFT vector **X** according to the new vector **y**;

  o Update $A^2 = \sum |\mathbf{y}|^2 / M$ and $\sigma^2 = A^2 M \dfrac{N-M}{N-1}$ ;

  o If $A^2 < \sigma_N^2$ **break** ;

**end for**

The algorithm uses DFT as a domain of sparsity but the same concept can be applied to other transform domains. This approach can provide successful signal reconstruction within a single iteration of the reconstruction algorithm. However, if the number of available samples *M* is very low, the iterative version of the algorithm is derived as well, updating the threshold value. If the inputs of the algorithm are vector of the *M* available samples **y**, signal length *N*, set of the available samples positions $\Omega_a=\{n_1,\ldots,n_M\}$, transform and measurement matrices $\mathfrak{I}$ and $\Phi$, CS matrix $\mathbf{A} = \Phi\mathfrak{I}$ and Gaussian noise variance $\sigma_N$, then the iterative version can be described using the Algorithm 5.

## 4. CS APPLICATIONS

The CS theory stating that the compressible signals can be efficiently reconstructed using a small set of incoherent measurements, motivated the researchers to explore possible fields of applications. Having in mind that many real-world signals satisfy the sparsity property, the applications ranges from the speech and audio signals [58]-[61], radar and communications [62]-[83], underwater, acoustic and linear frequency modulated signals [84]-[86], image reconstruction [87]-[98], biomedical applications [98]-[101], etc. The review of CS applications for different 1D signals, images and video data will be addressed in the sequel.

### 4.1. *CS devices (analog to information, single pixel camera, random lens imager)*

Let us firstly consider some of the hardware devices that are based on the CS principles.

(A) Duarte et al. in [102] proposed single pixel camera concept. This CS camera architecture is an optical computer, composed of a Digital Micromirror Device - DMD, two lenses and a single photon detector. It also contains an analog-to-digital (A/D) converter that computes random linear measurements of the scene under view. The image is recovered from the acquired measurements by a digital computer. Compared to the conventional silicon-based cameras, single pixel camera is a simpler, smaller, and cheaper and can operate efficiently across a much broader spectral range.

(B) Fergus et al. in [103] developed a random lens imaging technique. The technique uses a normal Digital single-lens reflex - DSLR camera, whose lens are replaced with a transparent material. The mirrors in this material are randomly distributed. The authors modified a Pentax stereo adapter in order to make one of the mirrors to have a random reflective surface. The CS measurements are in the form of images, obtained using this system. The new camera set-up has to be calibrated, in order to reconstruct the original image.

(C) Trakimas et al. in [104] proposed the design and implementation of an analog-to-information converter (AIC). The presented AIC is designed in a way that can sample at the Nyquist rate, but has also the CS operation mode. This design shows minimal complexity compared to conventional Nyquist rate sampling architectures. When dealing with signals with sparse frequency representation, this design has increased power efficiency of the sampling operation. To generate the pseudorandom sequence, a PN clock generator is used and it can be configured to provide a synchronous clock signal when Nyquist sampling is required.

## 4.2. *CS in biomedical applications*

CS finds usage in numerous biomedical applications, such as in Magnetic Resonance Imaging (MRI) [105]-[108], then electroencephalography (EEG), electrocardiography (ECG), electrooculography (EOG) and electromyography (EMG) signals, [109]-[120], etc. Some of the specific biomedical applications are given in the sequel.

(A)  Lowering the time of patient exposition to the harmful MR waves was the primary motivation of CS usage in MRI. However, the MR acquisition time is proportional to the dimensionality of the MR dataset, i.e., the number of spatial frequencies acquired. Scan time can be reduced by lowering the amount of data acquired, but still it has to be able to recover the whole information.

(B)  Lustig et al. in [98] implemented CS approach for rapid MRI imaging. Sparsity of the MRI in the transform domain is exploited for achieving two goals: reducing the scan time and improving the resolution of the observed fast spin-echo brain images and 3D contrast enhanced angiographs. The non-linear conjugate gradient solution is used for the optimization problem solving. The problem is in the form:

$$\arg\min_{\mathbf{x}} \|\mathfrak{I}_u \mathbf{x} - \mathbf{y}\|_2^2 + \chi \|\mathfrak{I}\mathbf{x}\|_1, \qquad (16)$$

where **x** is the image of interest, $\mathfrak{I}$ is an operator that transform signal from pixel representation into sparse representation, $\mathfrak{I}_u$ is an undersampled Fourier transform, **y** denotes measured *k*-space (e.g. frequency space) data from the scanner and $\chi$ is a regularization parameter. The conjugate gradient procedure is described in detail in [98].

(C)  Bioucas-Dias et al. introduced TwIST: Two-Step Iterative Shrinkage/ Thresholding Algorithms for Image Restoration [101]. This algorithm is introduced as an improved version of the Iterative Shrinkage Thresholding Algorithm (IST) – to overcome problem of its slow convergence in the cases when the measurement matrix **A** is ill-posed or ill-conditioned. The optimization problem is well-posed if has a solution, if a solution is unique and the solution changes continuously on the data [121]. Otherwise, the problem is ill-posed. If small perturbation of the **y** in the problem **y=AX** leads to large perturbation of the solution, the problem is ill-conditioned [121].

The algorithm is successfully applied on image deconvolution problems, as well as reconstruction of the images with missing samples. Considering the system of equations **y=AX**, the *t*-th iteration of the TwIST algorithm can be defined as follows:

$$\begin{aligned}\mathbf{X}_1 &= G_\eta(\mathbf{X}_0), \\ \mathbf{X}_{t+1} &= (1-\mu)\mathbf{X}_{t-1} + (\mu-\delta)\mathbf{X}_t + \delta G_\eta(\mathbf{X}_t),\end{aligned} \qquad (17)$$

where $\mu$ and $\delta$ are nonzero parameters. The starting value for the vector $\mathbf{X}$, $\mathbf{X}_0$, can be user-defined or $\mathbf{X}_0 = \mathbf{A}^{-1}\mathbf{y}$. Function $G_\eta$ is defined by using denoising operator $\Psi_\eta$ as:

$$G_\eta(\mathbf{X}) = \Psi_\eta\left(\mathbf{X} + \mathbf{A}^T(\mathbf{y} - \mathbf{A}\mathbf{X})\right), \tag{18}$$

where $\Psi_\eta(\varepsilon) = \arg\min_{\mathbf{X}} v^{-1}\Phi_{reg}(\mathbf{X}) + \|\mathbf{y} - \mathbf{A}\mathbf{X}\|^2/2$ and $\Phi_{reg}(\mathbf{X})$ is a regularization function.

The application of the TwIST algorithm in MRI reconstruction is shown. Fig. 1 shows an example of MRI reconstruction when only 2% of the image samples are available. The samples are acquired from the 2D DFT domain using a mask. The mask is formed of radial lines and placed around the origin. The TV regularization is done according to [101]. The original image, mask and the reconstructed image are shown in Fig. 1.

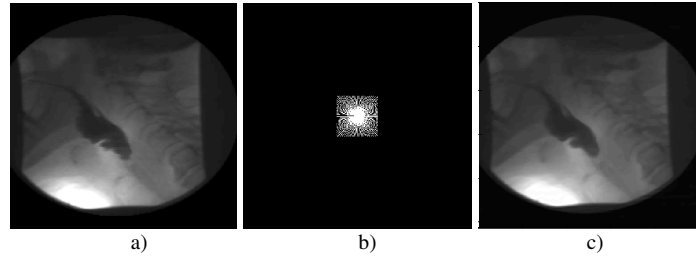

a)          b)          c)

Fig. 1 : a) Original image ; b) Mask in the 2D DFT domain ; c) Image reconstructed from available samples (2% of the total number of samples)

(D) Trzasko and Manduca in [122] proposed a method for under-sampled MR images recovering by using homotopic approximation of the $\ell_0$-norm. It is shown that the computed local minima of the homotopic $\ell_0$-minimization problem allows very highly undersampled $K$-space image reconstruction. The optimization problem can be defined starting from the relation:

$$u = \arg\min_{u} \|\mathbf{\Psi}u\|_0 \quad \text{subject to} \quad \mathbf{\Phi}u = \mathbf{\Phi}f, \tag{19}$$

where $\mathbf{\Psi}$ is wavelet, curvelet, etc. operator, $\mathbf{\Phi}$ is Fourier sampling operator and $f$ is the continuous signal. If the $\ell_0$ semi norm is replaced with the $\ell_1$ norm, as proposed by Candes and Donoho [122], the optimization problem can be recast as:

$$u = \arg\min_{u} \|\mathbf{\Psi}u\|_1 \quad \text{subject to} \quad \|\mathbf{\Phi}u - \mathbf{\Phi}f_n\|_2^2 \leq \varepsilon, \tag{20}$$

where measured data $f_n$ is noisy and $\varepsilon$ denotes the statistic of the noise process. Chartrand [122] proposed an alternative to the $\ell_0$ semi norm, that provides better sampling bounds compared to the $\ell_1$ and that is computationally feasible. He proposed usage of the $\ell_p$ semi norms ($0 < p < 1$].

The zero semi-norm of the signal can be defined as:

$$\|u\|_0 = \sum_{\Omega} \mathbf{1}(|u(n)| > 0), \tag{21}$$

with $\Omega$ that denotes the image domain and $\mathbf{1}$ is the indicator function. Any semimetric functional $\rho$ that satisfies following relation [122]:

$$\lim_{\sigma \to 0} \sum_{\Omega} \rho(|u(n)|, \sigma) = \sum_{\Omega} \mathbf{1}(|u(n)| > 0), \tag{22}$$

and if the σ is sufficiently small, can be used as a sparsity prior. Based on the previous relations, a new reconstruction paradigm can be defined as [122]:

$$\min_{u} \lim_{\sigma \to 0} \sum_{\Omega} \rho\left(|\Psi u(n)|, \sigma\right) \text{ subject to } \|\Phi u - \Phi f_n\|_2^2 \leq \varepsilon , \qquad (23)$$

where $\Psi$ is sparsifying basis and $\Phi$ is the CS matrix. The class of functionals that satisfy (22) are homotopic with $\ell_0$ semi norm. The proposed method is tested on the MR images, but the application to other medical images, such as the *x*-ray CT will be investigated in the future.

(E)  Abdulghani et al. in [111] analyzed performance of different reconstruction algorithms, applied for reconstruction of the EEG signals. They show that the best reconstruction results, with minimal error, are obtained by using the BP algorithm. This approach provides better reconstruction quality compared to the greedy approaches, but has greater computational cost. Six different EEG dictionaries have been observed and it is proved that the B-Spline dictionaries are the most suitable for CS of the EEG signals.

(F)  The approach proposed in [118]-[120] is used for the reconstruction of the under-sampled ECG signals, more precisely, the QRS complexes within the ECG signals. It is shown that the reconstruction of ECG signals can be done using just a few coefficients from the Hermite transform domain. The reconstruction is done by using gradient-based algorithm (Algorithm 2), where DFT is replaced by the Hermite transform.

A continuous signal $x(t)$ can be represented in terms of Hermite functions as follows:

$$x(t) = \sum_{p=0}^{\infty} c_p \Im_p(t) , \qquad (24)$$

where $c_p$ denotes Hermite expansion coefficients, while $\Im_p(t)$ denotes the *p*-th Hermite function. Signal $x(t)$ needs to be sampled at the non-uniform points that correspond to the roots of the Hermite polynomial or, uniformly sampled signal should be interpolated to obtain requested signal values. If **c** and **x** are Hermite coefficients and signal vector, respectively, and **H** denotes the transform matrix, then the signal expansion in the matrix form can be written as **c=Hx**:

$$\begin{bmatrix} c_0 \\ c_1 \\ \ldots \\ c_{P-1} \end{bmatrix} = \frac{1}{M} \begin{bmatrix} \Im_0(t_1)/(\Im_{M-1}(t_1))^2 & \ldots & \Im_0(t_M)/(\Im_{M-1}(t_M))^2 \\ \Im_1(t_1)/(\Im_{M-1}(t_1))^2 & \ldots & \Im_1(t_M)/(\Im_{M-1}(t_M))^2 \\ \ldots & \ldots & \ldots \\ \Im_{M-1}(t_1)/(\Im_{M-1}(t_1))^2 & \ldots & \Im_{M-1}(t_M)/(\Im_{M-1}(t_M))^2 \end{bmatrix} \begin{bmatrix} x(t_1) \\ x(t_2) \\ \ldots \\ x(t_M) \end{bmatrix} . \qquad (25)$$

To obtain the values at the non-uniform points, the signal sampled according to the sampling theorem is interpolated by using the *sinc* interpolation formula:

$$x(\lambda t_m) \approx \sum_{n=-K}^{K} x(nT) \frac{\sin\left(\pi(\lambda t_m - nT)/T\right)}{\pi(\lambda t_m - nT)/T}, \qquad (26)$$

where $m = 1,\ldots, M$, and $T$ is the sampling period. The time-axis scaling parameter $\lambda$ is introduced instead of $\sigma$, in order to avoid stretching and compressing of the basis functions. The parameter $\sigma$ is fixed ($\sigma=1$) and an optimal value of the parameter $\lambda$ (producing the best possible concentration - sparsity) is found according to the $l_1$-norm optimization:

$$\lambda_{opt} = \min_{\lambda} \|\tilde{\mathbf{c}}\|_1 = \min_{\lambda} \sum_{p=0}^{M-1} |\tilde{c}_p| = \min_{\lambda} \|HT\{x(\lambda t_m)\}\|_1 , \qquad (27)$$

where the operator HT{.} is used to denote the Hermite transform. The results obtained by applying the proposed procedure are shown in Fig. 2. The observed QRS complex is reconstructed by using only 55% of the available samples, with MSE of order $10^{-3}$.

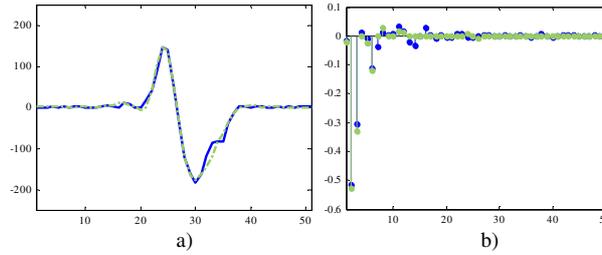

Fig. 2 : Reconstruction results for QRS complex: a) desired (solid line) and reconstructed signal (dashed line), b) Hermite coefficients of desired (blue) and reconstructed signal (green)

### *4.3. Applications in communications and radar signal processing*

The CS applications in radars [62]-[72] and communications [123]-[126] are widely studied in the literature. Hence, we describe some of these applications in wireless, ultra wideband (UWB) and Inverse synthetic aperture radar (ISAR) systems.

(A) Zhang et al. in [123] proposed CS application in UWB communications. CS is used to reduce the high data-rate of ADC at receiver. The receiver design is simpler with only one low-rate A/D. Here, the CS exploits the time sparsity of the signal through a filter-based CS approach applied on continuous time signals.

(B) Weiss in [125] used the CS in distributed radar network, composed of Wireless Local Area Network - WLAN routers as transmitters. The corresponding receivers are widely separated in the context of sparse modeling. The CS is applied in order to reduce the number of samples transferred to the central processing stage and it is used for the estimation of multiple targets positions and velocities. The paper uses IEEE 802.11b signal. The proposed approach is compared with the matched filter technique that is the traditional approach for target determination.

(C) Particular interest of using the CS approach in communications raised for signals belonging to the two interfering standards in wireless communications: Bluetooth and IEEE 802.11b that share Industrial, Scientific and Medical (ISM) frequency band [75],[82]. The first standard uses frequency hopping spread spectrum modulated signals (FHSS), while the second standard applies direct sequence spread spectrum (DSSS) modulation. Different sparsification basis are considered for different signals. Original signal, consisted of FHSS and DSSS components, does not satisfy sparsity property, which is not the case with the separated components. Therefore, the first step is components separation, done by using an eigenvalue decomposition method. The EVD of the covariance matrix **C** is defined as:

$$\mathbf{C} = \mathbf{U}\mathbf{\Lambda}\mathbf{U}^{\mathbf{T}} = \sum_{i=1}^{N+1} \lambda_i u_i(n) u^*_i(n), \tag{28}$$

where **U** is an eigenvectors matrix, **Λ** is a diagonal eigenvalues matrix where eigenvalues are sorted in decreasing order, $\lambda_i$ are eigenvalues and $u_i$ are eigenvectors. The covariance matrix is defined based on the TF representation – the inverse form of the S-method is used for definition of the covariance matrix:

$$C_K(n+m, n-m) = \frac{1}{N+1} \sum_{k=-N/2}^{N/2} SM(n,k) e^{j\frac{4\pi}{N+1}mk}. \tag{29}$$

The EVD is applied to the matrix $\mathbf{C_K}$ (where $K$ denotes the number of signal components) according to the relation (28), resulting in the eigenvectors that correspond to the signal components. The choice of the sparsity domain is made between the DFT and HT domain, by measuring the concentration using the $\ell_1$-norm and choosing the one producing the minimal concentration.

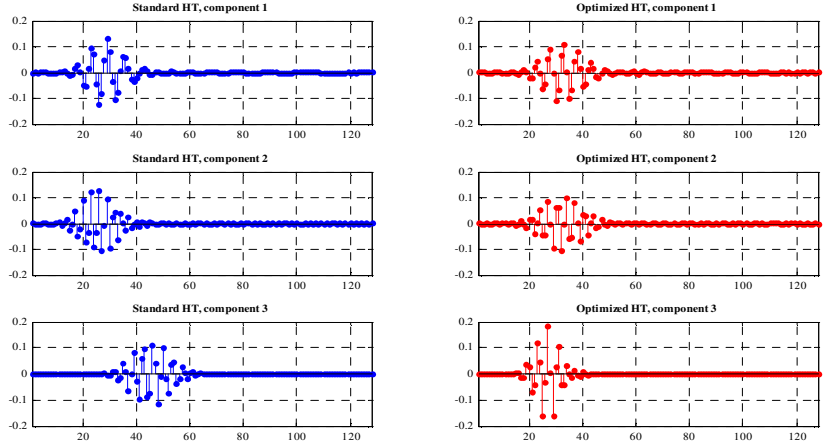

Fig. 3 HT and the optimized HT of the first 3 eigenvectors that correspond to the FHSS signal components

After sparsification domain is chosen, the eigenvectors are randomly under-sampled and reconstructed from only 50% of the acquired samples. Under-sampled vectors are reconstructed by using the $\ell_1$-norm minimization according to (9).

The reconstructed components from the 50 % of measurements are shown in Fig. 4 (blue denotes the original while red is the reconstructed component). It is shown that the HT is chosen as a sparsifying basis for the FHSS, while the DFT domain is used for the DSSS components. In the case of HT, the eigenvectors should be resampled at non-uniform points being proportional to the roots of the Hermite polynomial [127].

Signal having 3 FHSS and 4 DSSS components is observed. The first 3 eigenvectors correspond to the FHSS components and therefore the optimized HT is calculated (Fig. 3). It can be seen that the optimized HT provides more compact representation compared to the standard HT. The rest of the eigenvectors correspond to the DSSS components and the DFT is used as a sparsification basis in this case.

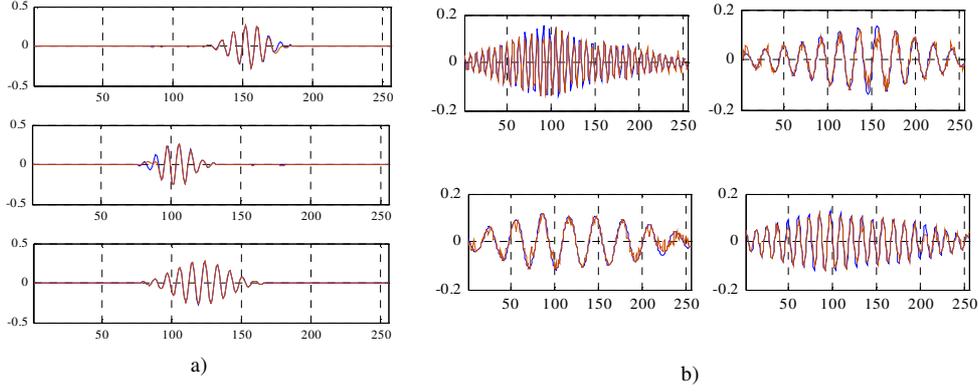

Fig. 4 Separated and reconstructed components of the a) FHSS and b) IEEE 802.11b signals. Blue is the original component, brown is the CS reconstructed component. The percent of the randomly selected samples is 50%

(D) Lu et al. in [128] proposed a novel Distributed Streaming Compressive Spectrum Sensing (DSCSS) algorithm. This approach is used for wideband spectrum sensing under decentralized cognitive radio network (CRN) scenario. The algorithm has low computational complexity that makes it suitable for the on-line applications. DSCSS uses an analog-to-information (AIC) convertor in acquisition process. An AIC showed to be suitable for streaming framework. A sliding time-window of length equal to NT is applied to the sensed signal, while $M$ sub-Nyquist samples are produced according to one time-windowed signal.

DSCSS approach does not require sparsity of the signal to be a priori known. In practical applications, sparsity of the wide-band spectrum is not available. Therefore, DSCSS estimates sparsity and the support set of the spectrum. This support set is then exchanged within the cognitive radio network as an a priori information, in order to obtain a cooperative sensing gain. This is done with a goal to overcome wireless fading effects.

(E) The CS application in recovery of the ISAR signal rigid body, by exploiting the concept of sparsity, is proposed in [58], [72]. The ISAR signal contains both, the rigid body and micro-Doppler segment caused by fast-moving parts of a target. The observed segments of the signal partially overlap in the time-frequency plane. The method finds and removes the overlapping values and recovers a rigid body signal. The separation of the rigid body stationary and the micro-Doppler nonstationary signal segments is done by using the sorted Short-Time Fourier transform (STFT) values along the time axis. Then a certain percent of the strongest STFT values, for each frequency, is removed, which results in reduction (or elimination) of the micro-Doppler nonstationary components. The remaining part of the STFT will contain the rigid body components.

A radar return signal after coherent processing and filtering is consisted of two components, rigid body $x_r(t)$ and micro-Doppler $x_m(t)$:

$$x(t) = x_r(t) + x_m(t), \quad x_r(t) = \sum_{i=1}^{K} r_i e^{j\frac{2\pi f_{oi} t}{N}}, \quad (30)$$

where $r_i$ denotes components' amplitudes and $f_{oi}$ are components' frequencies. The STFT is used for the TF representation of $x(t)$:

$$S(t,f) = \sum_{l=0}^{L-1} x(t+l) e^{-j\frac{2\pi l f}{L}}, \text{ or } \mathbf{S}_L(t) = \mathbf{E}_L \mathbf{x}(t), \quad (31)$$

where

$$\mathbf{S}_L(t) = [S(t,0),...,S(t,L-1)]^T, \quad \mathbf{x}(t) = [x(t), x(t+1),..., x(t+M-1)]^T,$$
$$E(l,f) = e^{-j\frac{2\pi f l}{L}}, \quad (32)$$

and $\mathbf{E}_L$ is the $L \times L$ DFT matrix. Non-overlapping STFT is used in TF analysis, calculated with step $L$ in time $t$, and we can write the following relations:

$$\mathbf{S} = [\mathbf{S}_L(0)^T, \mathbf{S}_L(M)^T, ..., \mathbf{S}_L(N-M)^T]^T = \mathbf{E}_{L,N} \mathbf{x}. \quad (33)$$

Note that the $S(t,f)$ denotes scalar STFT value at time $t$ and frequency $f$, $\mathbf{S}_L(t)$ denotes STFT vector that contains $L$ frequencies at instant $t$, while $\mathbf{S}$ is a vector of all STFT values for all frequencies $f$ and all time instants $t$. The $N \times N$ matrix $\mathbf{E}_{L,N}$ and the signal vector $\mathbf{x}$ are formed as:

$$\mathbf{E}_{L,N} = \begin{bmatrix} \mathbf{E}_M & \mathbf{0}_M & ... & \mathbf{0}_M \\ \mathbf{0}_M & \mathbf{E}_M & ... & \mathbf{0}_M \\ ... & ... & ... & ... \\ \mathbf{0}_M & \mathbf{0}_M & ... & \mathbf{E}_M \end{bmatrix}, \quad (34)$$
$$\mathbf{x} = [\mathbf{x}(0)^T, \mathbf{x}(M)^T, ..., \mathbf{x}(N-M)^T] = [x(0), x(1), ..., x(N-1)]^T.$$

Vector $\mathbf{x}$ can be written using the inverse DFT matrix $\mathbf{E}_N^{-1}$ and DFT vector $\mathbf{X}$, and following relations are obtained:

$$\mathbf{x} = \mathbf{E}_N^{-1} \mathbf{X}, \quad \mathbf{S} = \mathbf{E}_{L,N} \mathbf{x} = \mathbf{E}_{L,N} \mathbf{E}_N^{-1} \mathbf{X} = \mathbf{\Psi} \mathbf{X}, \quad (35)$$

where $\mathbf{\Psi}$ is the $N \times N$ transformation matrix. After calculating the STFT according to (35), the sorting operation along is performed along the frequency (for each $f$):

$$s_{sort}(t,f) = sort\{S(t,f)\}, \quad (36)$$

where $t=0, ..., L-1$. The sorting operation is performed in ascending order and certain percentage $P$ of low value coefficients and percentage $Q$ of high value coefficients are removed from the sorted STFT:

$$s_{cs}(f) = \{S_{SORT}(n,f), n = P, P+1, ..., L-Q\} \quad (37)$$

where $s_{cs}(f)$ denotes the vector of available STFT coefficients at frequency $f$, while $\mathbf{S}_{CS}$ denotes vector for all frequencies. The vector of all available STFT points is $\mathbf{S}_{CS} = \mathbf{\Psi}_{CS} \mathbf{X}$, where the matrix $\mathbf{\Psi}_{CS}$ is formed by omitting the rows which corresponds to the removed STFT values. The CS problem can be formulated as follows:

$$\min \|\mathbf{X}\|_{\ell_1} \text{ subject to } \mathbf{S}_{CS} = \mathbf{\Psi}_{CS} \mathbf{X}. \quad (38)$$

The results of the procedure applied on real radar signal, are shown in Fig. 5. The signal consists of the rigid body and three corner reflectors rotating at ∼ 60 RPM. The 80% of the samples from the sorted STFT are omitted and rigid body is reconstructed from the rest of the samples.

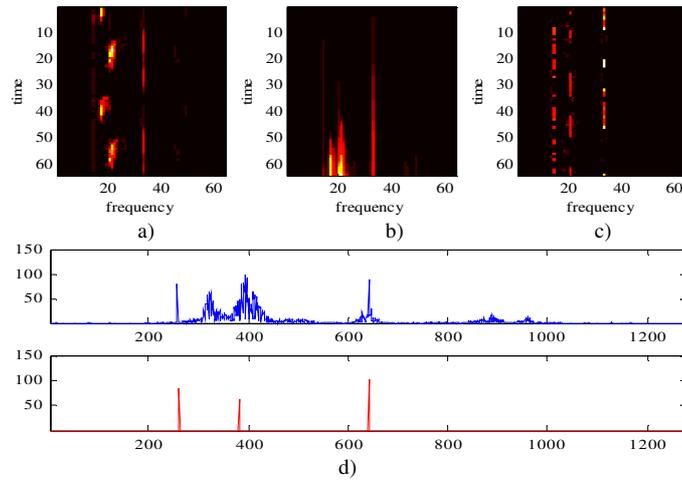

Fig. 5 : Real radar signal : a) STFT, b) sorted STFT, c) STFT that remains after discarding certain region from the sorted STFT, d) the original DFT transform - blue, and the reconstructed DFT.

(F)    The same approach is used in [75] for separation of signals belonging to two different interfering wireless standards - Bluetooth and IEEE 802.11b standard. The proposed procedure works even in the case of overlapping signal components. Based on the a priori knowledge on the signals' nature, it is possible to select and extract a small set of time-frequency points that entirely belongs to the IEEE 802.11b signal. The extracted points are used to recover the full signal by using the described CS-based approach.

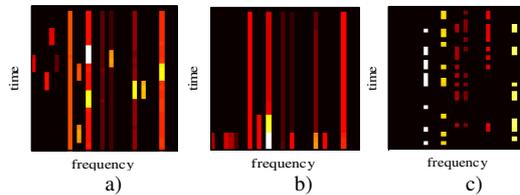

Fig. 6 a) STFT of the original signal, b) sorted STFT, c) remaining STFT values after discarding certain region from the sorted STFT

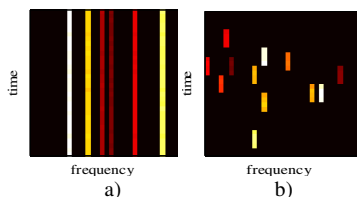

Fig. 7 : Reconstructed STFT of : a) IEEE 802.11b, b) Bluetooth signal

Once the components of IEEE 802.11b signal are extracted, the remaining components in the time-frequency plane belong to the Bluetooth signal. The example of signal consisted of Bluetooth and IEEE 802.11b components that overlap, is shown in Fig. 6. Six components belong to the IEEE 802.11b signal, while twelve components (of short duration) belong to the Bluetooth signal. The STFT is sorted and certain percent of the lowest and highest energy samples are removed (sorted and remained STFT are shown in Fig. 6b and 5c). The recovered STFT is shown in Fig. 7a and corresponds to the IEEE 802.11b signal, while the remaining part of the TF coefficients correspond to the Bluetooth signal.

(G)    Zhang et al. in [76] presented a high-resolution inverse synthetic aperture radar imaging and showed that the proposed CS imaging outperforms the conventional range-Doppler approach, regarding the image resolution. More details regarding the proposed procedure follow.

If the signal of interest is defined as:

$$x(t) = \sum_{k=1}^{K} E_k \rho(t/T_a) e^{-j2\pi f_k t} + \gamma(t), \qquad (39)$$

where $K$ denotes the number of the strongest scattering centers, $\gamma(t)$ is the synthetic additive noise in the range cell, $E_k$ is the scattering amplitude, $\rho(t/T_a)$ is the unit rectangular function, $f_k$ is the carrier frequency and $T_a$ is the time width of chirp pulse. If we denote the transform matrix as $\Im$ and sensing matrix as $\mathbf{A}$, then the previous relation can be rewritten as:

$$x(t) = \Im\theta + \gamma(t), \qquad (40)$$

where $\theta$ is the vector whose non-zero components correspond to the complex amplitudes of the $K$ strongest scattering centers. The optimization problem is recast to:

$$\min \|\hat{\theta}\|_1 \quad \text{subject to} \quad \|x(t) - \mathbf{A}\Im\hat{\theta}\|_2 \leq \xi, \qquad (41)$$

where $\xi$ is the noise level, and measurement vector $\mathbf{y}$ is: $\mathbf{y}=\mathbf{Ax}$ with $\mathbf{A}_{M \times N}$ ($M<N$) measurement matrix. Here, the assumption is that the precise motion compensation is done. It is not always the case in large-size maneuvering targets, which is one of the drawbacks of the proposed method.

### 4.4.    *Compressive sensing image reconstruction*

(A)    Musić et al. in [87] proposed a method that combines CS based image reconstruction and object detection algorithm. The method is used in search and rescue application and allows image recovering if up to 80% of pixels is unavailable or corrupted by noise. Preserving good quality of the reconstructed image is of particular importance for object detection algorithm. The solution is based on the 2D DCT domain and gradient descent recovery method.

Through the iterations, a two version of image $I$ are formed: $I^+(n,m)$ and $I^-(n,m)$, where $(n,m)$ denotes pixels positions:

$$\begin{aligned} I^+(n,m) &= I_i(n,m) + \Delta\delta(n-n_k, m-m_k) \\ I^-(n,m) &= I_i(n,m) - \Delta\delta(n-n_k, m-m_k) \end{aligned}, \qquad (42)$$

where the correction factor $\Delta$ is $\Delta=\text{mean}\{I(n,m)\}$ and $\delta$ is 2D discrete delta function. The next step is gradient vector calculation:

$$G_i(n_k, m_k) = \begin{cases} \dfrac{1}{N}\left\{\left\|\aleph(I^+)\right\|_1 - \left\|\aleph(I^-)\right\|_1\right\}, & \text{for } (n_k, m_k) \in \Theta \\ 0, & \text{otherwise} \end{cases}, \qquad (43)$$

where $N$ is the total number of pixels, $\aleph$ denotes 2D DCT and $\Theta$ is the set of positions corresponding to the missing pixels: $\Theta=\{(n_k,m_k),\ k=1,\ldots,N_M\}$, $N_M$-number of missing pixels. In each iteration, the image is updated: $I_{i+1}(n,m) = I_i(n,m) - G_i(n,m)$. By minimizing the $\ell_1$-norm of the sparsity measure the missing samples are changed toward the exact values.

Finding a potential targets on the CS-reconstructed image is done by observing parts of the image. The image preprocessing assumes conversion from the RGB to $YC_bC_r$ color model. The image is firstly divided into the set of $C$ non-overlapping clusters denoted as $K$, using means shift clustering algorithm. The clusters correspond to significant image features, i.e. dominant colors. If an image $I$ consists of set of clusters $K$, then:

$$I = \cup_{c=1}^{C} K_j\ . \qquad (44)$$

Clusters are further divided into the sets of segments $S_{j,k}$. One segment represents a spatially connected component region. If the total number of segments in a cluster is denoted as $Q$, then the cluster can be defined as:

$$K_j = \cup_{k=1}^{Q} S_{j,k}\ . \qquad (45)$$

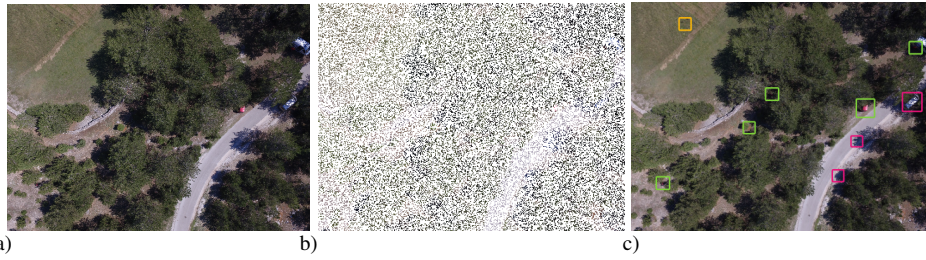

a)  b)  c)
Fig. 8 : Reconstruction results for search-and-rescue image. 70% of pixels at random positions are missing: a) Original image; b) Image with missing samples; c) Reconstructed image

The segment $S_{j,k}$ will be considered as a target if the following conditions are satisfied:

$$s(K_j) < T_1\ \wedge\ s(S_{j,k}) > T_2\ \wedge\ Q \leq N_A\ , \qquad (46)$$

where $s$ denotes size of the set, $T_1$ and $T_2$ are threshold values and $N_A$ is the maximum allowed number of candidate segments in a cluster.

The algorithm's performance in case of 70% of missing samples is shown in Fig. 8. The reconstruction is done on 16 × 16 blocks in 50 iterations. Green squares represent correct detections (objects found in both, original and reconstructed images), red squares represent object found on the original, but not found on the reconstructed image and orange squares represent object found on the reconstructed but not found on the original image.

(B) An application of the adaptive threshold based algorithm for the ISAR images reconstruction is done within the paper [96]. The algorithm, initially proposed in [98] for 1D data, is adapted for the 2D data and exhibits sparsity in 2D

DFT domain. It is based on the analytically derived threshold that precisely separates signal and non-signal components. The approach provides efficient results even if less than 10% of the samples are available.

If we denote the full and partial sets of signal samples as $S_F$ and $S_P$:

$$S_F = \{S_F(x, y): x \in \{1,...,I\}, y \in \{1,...,J\}\}, \\ S_P = \{S_P(x, y): x \in \{1,...,M\}, y \in \{1,...,L\}\}, \quad S_P \subset S_F, \tag{47}$$

where $I \times J$ denotes the total number of samples, $M$ and $L$ number of samples along $x$ and $y$ directions respectively ($M<I$, $L<J$), the variance of noise, that appears as a consequence of missing signal samples, in 2D case can be calculated as follows:

$$\sigma^2 = \sum_{i=1}^{K} E_i^2 ML \frac{IJ - ML}{IJ - 1}, \tag{48}$$

where $E_i$ denotes amplitude of the $i$-th component. The above relation is valid for $K$ signal components. If the probability that all DFT values at noisy components positions are below the signal components is $P=0.99$, then the threshold that separates signal and non-signal components in the 2D DFT domain can be calculated as follows:

$$T = \sigma \sqrt{-\log(1 - P^{1/(IJ)})}. \tag{49}$$

---

**Algorithm 7: Iterative SFAR-2D algorithm**

**Input:** vector of the available samples $\mathbf{y}$, set of the available samples positions $\Omega_a$, 2D DFT and inverse 2D DFT matrices $\Im$ and $\Psi$, and $\mathbf{k} = \emptyset$

- Calculate variance: $\sigma^2 = \sum_{i=1}^{K} E_i^2 ML \frac{IJ - ML}{IJ - 1}$, $E^2 = \sum_{i=1}^{K} E_i^2$;
- Calculate threshold $T$ for a given $P=0.99$;
- Calculate the initial DFT vector $\mathbf{Y}$ that corresponds to $\mathbf{y}$: $\mathbf{Y} = \Im_{\Omega_a} \mathbf{y}$;
- **for** $i=1 : i=i+1$: **until** all components are detected
  $\mathbf{k} = \mathbf{k} \cup \arg\{|\mathbf{Y}| > T\}$;
  Calculate $\mathbf{F} = (\mathbf{A}^H \mathbf{A})^{-1} \mathbf{A} \mathbf{y}$, $\mathbf{A} = \Psi(\Omega_a, \mathbf{k})$;
  Update $\mathbf{y}$, $\mathbf{Y}$, $E^2$ and $\sigma^2$: $\mathbf{y}=\mathbf{y}-\mathbf{A}\mathbf{Y}(\mathbf{k})$; $\mathbf{Y} = \Im_{\Omega_a} \mathbf{y}$; $E^2 = \sum (|\mathbf{y}|^2 / (ML))$ and
  $\sigma^2 = \sum_{i=1}^{K} E_i^2 ML \frac{IJ - ML}{IJ - 1}$;
  **end for**

**Output:** Reconstructed DFT $\mathbf{F}$

---

Assume that the set of the available samples positions is denoted as $\Omega_a$, $\Omega_a = \{(x_1, y_1),...,(x_M, y_L)\}$, $\Psi$ denotes inverse 2D DFT, $\mathbf{y}$ is vector of measurements and $\mathbf{k}$ denotes set of positions. The vector of initial DFT is calculated as:

$$\mathbf{Y} = \Im_{\Omega_a} \mathbf{y}, \tag{50}$$

where $\Im$ denotes 2D DFT matrix obtained as a Kronecker product of two $I \times J$ DFT matrices: $\Im = DFT \otimes DFT$ and matrix $\Im_{\Omega_a}$ contains all rows of the full 2D DFT matrix $\Im$ and only those columns defined by the set $\Omega_a$.

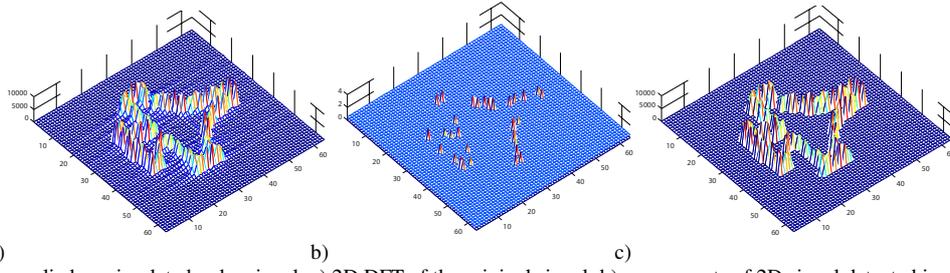

a)            b)            c)

Fig. 9 : Procedure applied on simulated radar signal: a) 2D DFT of the original signal; b) components of 2D signal detected in a single step, c) 2D DFT of the reconstructed signal

The iterative version of the algorithm is summarized through the Algorithm 7 (Simple and Fast Algorithm for Reconstruction of 2D signals - SFAR 2D [96]). The algorithm is tested on the simulated radar signal. It is assumed that only 25% of the samples are available. The 2D DFT of the original signal is shown in Fig. 9a. The components of the initial DFT, selected by the threshold, and the reconstructed 2D DFT of the signal, are shown in Fig. 9b and c, respectively.

(C) Wu et al. proposed method for Multivariate Compressive Sensing for Image Reconstruction in the Wavelet Domain - multivariate pursuit algorithm (MPA) [130]. Statistical structures of the wavelet coefficients are exploited, having in mind significantly statistical dependency that exists among the wavelet coefficients of images. The observed images are sparse or compressive in wavelet domain.

**Algorithm 8: MPA**

**Input**: CS matrix $\mathbf{A}_{K \times M'}$, measurement matrix $\mathbf{\Omega}$.

**Output**: Approximation $\hat{\mathbf{X}}$ to the true coefficients matrix $\mathbf{X}$

1. Set $n=0$, $\hat{\mathbf{X}} = 0$, $\mathbf{rY}^0 = \mathbf{Y}$

2. Set $\hat{\mathbf{X}}^n = 0$, $\mathbf{r}^0 = \mathbf{rY}^n$. Compute the vector of residual correlations $\mathbf{c}$: $\mathbf{c} = \mathbf{A}^T \mathbf{r}^0$

3. Form the index set $\mathbf{I}$:
   $\mathbf{I} = \{I(1), ..., I(M')\}$,
   $\left| \sum_{j=1}^{q} \left| c_{I(1)}^{j} \right| \right| > ... > \left| \sum_{j=1}^{q} \left| c_{I(M')}^{j} \right| \right|$ ;

4. Estimate successively the components of $\mathbf{X}^n$ according to $\mathbf{I}$:
   a) $i=1$, $\mathbf{r}^i = \mathbf{r}^0$;
   b) Solve the following sub-problem to estimate $\bar{\mathbf{x}}_{I(i)}^n$: $\mathbf{r}^i = \mathbf{a}_{I(i)} \left( \bar{\mathbf{x}}_{I(i)}^n \right)^T + \boldsymbol{\varepsilon}^i$ ;
   c) Update the residual: $\mathbf{r}^{i+1} = \mathbf{r}^i - \mathbf{a}_{I(i)} \left( \bar{\mathbf{x}}_{I(i)}^n \right)^T$, $\mathbf{a}_{I(i)}$ - $I(i)$-th column of $\mathbf{A}$;
   d) $i=i+1$; if $i \leq M'$ go to step b);

5. Update $\hat{\mathbf{X}} = \hat{\mathbf{X}} + \hat{\mathbf{X}}^n$, $\mathbf{rY}^{n+1} = \mathbf{r}^{M'}$ ;

6. $n=n+1$ and go to the step 2. Stop if $n=N+1$ or $\left( \sum_{k=1}^{K} \sum_{j=1}^{q} \left| \mathbf{rY}_{k,j}^{n+1} \right|^2 \right) / (Kq) < \varepsilon_0$ .

Unlike the traditional CS where the measurements are taken by firstly rearranging the coefficients of images to a vector and then projecting it with sensing matrix, here the coefficients are rearranged according to the partitioned neighborhoods.

The CS measurements consist of two parts: the scaling coefficients measured directly, and the compressed samples of the wavelet coefficients, obtained by projecting the wavelet coefficients with the sensing matrix. If we denote the CS matrix as $\mathbf{A}_{K \times M'}$, measurement matrix $\mathbf{\Omega}$, $M' = \lceil M/q \rceil$, where $M$ is the total number of wavelet coefficients and $\lceil . \rceil$ denotes ceiling function (case $q=1$ corresponds to the coefficients vector of traditional CS), then the vector of multivariate measurements $\mathbf{Y}_{K \times q}$ ($K << M'$) is:

$$\mathbf{Y} = \mathbf{A}\mathbf{X}, \qquad (51)$$

where $K \times q$ is the number of measurements. Parameter $q$ denotes the size of the wavelet neighborhood. The Multivariate Pursuit Algorithm – MPA steps can be summarized through the Algorithm 8. By exploiting the statistical dependences among wavelet coefficients in multivariate algorithms, the reconstruction performance is much improved compared to the state-of-the art methods. The multivariate algorithms also have higher computational efficiency.

(D)    Bobin et al. in [131], [132] exploited the CS approach in astronomical images and astronomical instrument design. There is a growing interest in astronomical data compression, having in mind that the conventional data compression cannot be used in many cases. The system that encodes into the analog domain can be designed with an optical system that directly measures incoherent projections of the input image. There are three main properties that have to be under control: resolution (point spread function), sensitivity (ability to detect low level signals) and photometry.

The aim is to recover the original signal $\mathbf{x}$ from the compressible signal $\mathbf{y} = \mathbf{\Omega}\mathbf{x}$. If the measurement matrix is denoted $\mathbf{\Omega}$, transform matrix is denoted as $\mathbf{\Psi}$ denotes the transform matrix, and $\mathbf{X} = \mathbf{\Psi}\mathbf{x}$, then the optimal $\mathbf{X}$ can be found as a solution of:

$$\min_{\mathbf{X}} \|\mathbf{X}\|_1 \text{ subject to } \mathbf{y} = \mathbf{\Omega}\mathbf{\Psi}\mathbf{X}. \qquad (52)$$

More realistic case is when observations are corrupted by noise: $\mathbf{y} = \mathbf{\Omega}(\mathbf{x}+\mathbf{n})$, where $\mathbf{n}$ is a white Gaussian noise with variance $\sigma_n^2$. The projected noise $\mathbf{n}_\Omega = \mathbf{\Omega}\mathbf{n}$ is still white Gaussian noise with the same variance. The optimization problem can now be recast as:

$$\min_{\mathbf{X}} \|\mathbf{X}\|_1 \text{ subject to } \|\mathbf{y} - \mathbf{\Omega}\mathbf{\Psi}\mathbf{X}\|_2 \leq e, \qquad (53)$$

where $e$ denotes an upper bound of $\|n\|_2$. Suppose that $N$ observations of the same sky area are available $\mathbf{y}_i$, $i=1,\ldots,N$:

$$\mathbf{y}_i = \mathbf{\Omega}_{\lambda_i}\mathbf{\Psi}\mathbf{X} + \mathbf{n}_i, \quad \forall i = 1,\ldots,N, \qquad (54)$$

and $\mathbf{\Omega}_{\lambda_i}$, $i=1,\ldots,N$ are $N$ independent random submatrices of $\mathbf{\Omega}$, card($\lambda_i$)=$M$. Recovering $\mathbf{x}$ from $N$ compressed observations, i.e. the decompression problem (53) can now be defined as:

$$\min_{\mathbf{X}} \|\mathbf{X}\|_1 \text{ subject to } \sum_{i=1}^{N} \|\mathbf{y}_i - \mathbf{\Omega}_{\lambda_i} \mathbf{\Psi X}\|_2 \leq \sqrt{N} e, \quad (55)$$

Or can be recast into the Lagrangian form:

$$\min_{\mathbf{X}} \mu \|\mathbf{X}\|_1 + \frac{1}{2} \sum_{i=1}^{N} \|\mathbf{y}_i - \mathbf{\Omega}_{\lambda_i} \mathbf{\Psi X}\|_2^2. \quad (56)$$

That can be solved by using projected Landweber iterative algorithm [131]. The choice of the parameter $\mu$ is important issue, due to its role as a balance between sparsity constraint and data fit of the solution. More details can be found in [131].

*4.5. Compressive sensing in video data*

(A) Beside the application in still image recovering, there are various situations when we need to apply the CS in video data [133]. However, the CS application in video has higher demands in terms of the complexity of imaging architectures, as well as reconstruction algorithms, in compare with the still-image CS [133]-[135]. Also, it can happen that a compressive camera does not capture a sufficient number of measurements to recover the frames of the video. Video recovery problems have high memory requirements and algorithm implementations require large dense matrix systems. Therefore, it requires high-performance hardware and fast iterative algorithms, in order to provide sufficiently high throughout.

Several compressive imaging architectures are designed:
- Spatial multiplexing cameras - apply CS multiplexing in space in order to improve the spatial resolution of videos obtained from sensor arrays whose spatial resolution is low;
- Temporal multiplexing cameras - apply CS multiplexing in time with an aim to improve the temporal resolution of videos obtained from sensor arrays with low temporal resolution;
- Spectral and angular multiplexing cameras - apply CS multiplexing to sense variations of light in a scene beyond the spatial and temporal dimensions.

Several methods for CS video recovery are developed, among them, variational (constrained and unconstrained) and greedy methods.

Constrained variational approach deals with the problems in the form:

$$\hat{\mathbf{x}} = \arg\min_{\mathbf{x},\mathbf{X}} f(\mathbf{\Omega}, \mathbf{x}, \mathbf{y}) + g(\mathbf{X}) \text{ subject to } \mathbf{X} = \mathbf{\Psi x}. \quad (57)$$

Function $f$ is used to model the video acquisition process – optics, modulation and sampling, while the function $g$ is a regularization function. One example of the $f$ and $g$ functions definitions is in the case of frame-by-frame recovery by using 2D wavelet transform:

$$f(\mathbf{\Omega}, \mathbf{x}, \mathbf{y}) = \|\mathbf{y} - \mathbf{\Omega x}\|_2^2, \quad g(\mathbf{X}) = \mu \|\mathbf{X}\|_1, \quad (58)$$

where x denotes vectorized image frame, $\Omega$ is the sensing matrix, $\Psi$ is the 2D wavelet transform matrix and $\mu>0$ is a regularization parameter.

If we are dealing with an invertible transform $\Psi$ then the problem (57) can be defined as follows:

$$\widehat{\mathbf{X}} = \arg\min_{\mathbf{X}} f(\mathbf{\Omega\Psi^{-1}}, \mathbf{X}, \mathbf{y}) + g(\mathbf{X}) \qquad (59)$$

where $\mathbf{X}$ denotes the video or single video frame in the transform domain.

Greedy algorithms are used for the unconstrained problems. They are based on iterative constructing a sparse set of non-zero transform coefficients and finding solution of the minimization problem $\left\|\mathbf{\Omega\Psi^{-1}X-y}\right\|_2^2$. The minimization problem solution can be found using the following greedy algorithms: OMP, regularized OMP (ROMP) and stagewise OMP (StOMP), CoSaMP [133].

(B)     A new approach for estimation of the motion parameters in compressive sensed video sequences under a reduced number of randomly chosen video frames, is proposed in [134]. The method focuses on the velocity estimation and combines sparse reconstruction algorithms with time-frequency analysis, applied to µ-propagation signal. The µ-propagation maps the video frames sequence into the frequency modulated signal, or into the high nonlinear phase signal. If a video frame at the instant *t* is defined as:

$$F(x,y,t) = p(x,y) + o(\Delta x, \Delta y) \,, \qquad (60)$$

where $\Delta x = x-x_0-b_x t$, $\Delta y = y-y_0-b_y t$, $o(x,y)$ denotes the moving object, $p$ is background, $(x_0,y_0)$ denote an initial object position and $(b_x,b_y)$ is the velocity. The projection of the frame onto the *x*-axis is defined as:

$$R(x,t) = \sum_y F(x,y) = \sum_y p(x,y) + \sum_y o(\Delta x, \Delta y) = P(x) + O(\Delta x) \,. \qquad (61)$$

Finding derivative of the $R(x,t)$ with respect to $t$ and assuming the constant background, the following signal is obtained:

$$\widehat{R}(\Delta x) = \frac{\partial R(x,t)}{\partial t} = b_x \frac{\partial O(\Delta x)}{\partial x} \approx R(x,t-1) - R(x,t) \,. \qquad (62)$$

The velocity estimation is done by applying the TF analysis to the signal in the form:

$$m(t) = \sum_x \widehat{R}(\Delta x) e^{j\mu x} \,, \qquad (63)$$

having in mind that the instantaneous frequency corresponds to the moving object velocity. As TF representation, the S-method can be used since it provides cross-terms free representation and is more suitable in the noisy signal cases. It is defined based on the STFT as:

$$S_M(t,f) = \sum_{i=-L}^{L} STFT(t, f+j\boldsymbol{\theta}) STFT^*(t, f-j\boldsymbol{\theta}), \qquad (64)$$

where $L$ is the S-method window width, while the STFT($t,f$) is defined as the FT of the windowed signal $m(t)$, with window function $w(\tau)$: $STFT(t,f) = \sum_\tau w(\tau) m(t+\tau) e^{-j\omega\tau}$ .

The CS is employed to reduce number of frames required for the IF estimation. In other words, the CS is used to assure motion parameters estimation from an incomplete set of frames. If the subset of frame is denoted as $S$, $S(x,y,t_s) \subset F(x,y,t)$, where only $M$ frames are acquired, $t_s=\{t_1,\ldots,t_M\}$, then the μ propagation vector will contain small number of samples, i.e. we will have signal $m(t_s)$. For each windowed signal part used for the STFT calculation, we have the measurement vector $y(t_{si})$:

$$y(t_{si}) = w(\tau)m(t_{si} + \tau), \quad \forall t_{si} \in t_s, \tag{65}$$

instead of desired vector $x(t)=w(\tau)m(t+\tau)$. The FT of the vector $y(t_{si})$ will produce low resolution in the STFT, and therefore, the CS is used in this step to recover missing samples in the vector $y(t_{si})$ and improve the resolution.

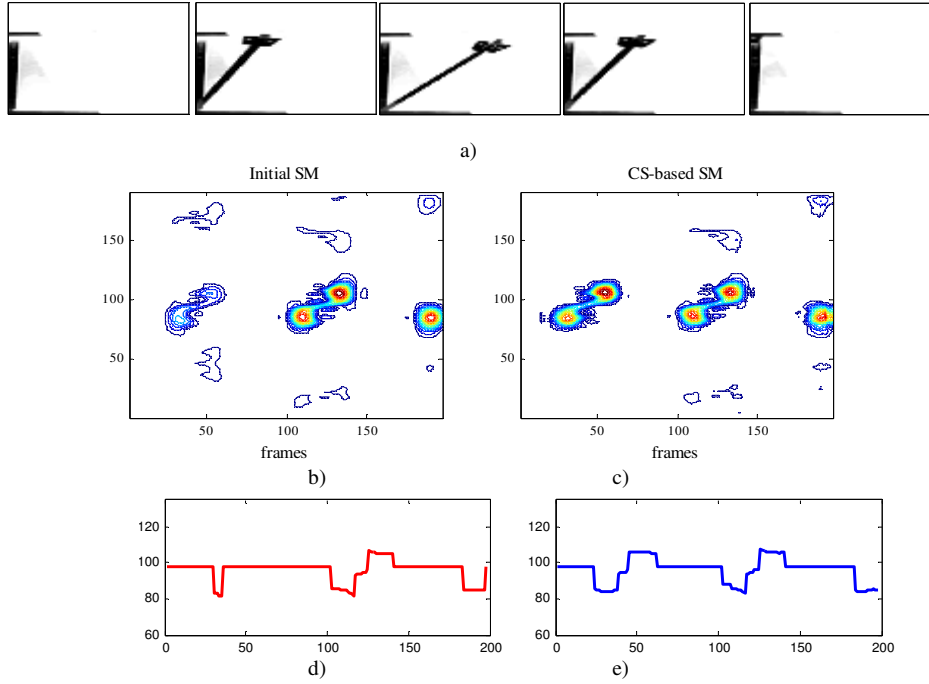

Fig. 10 : a) Several frames from the observed video sequence; b) Initial S-method of variable μ-propagation vector; c) CS based S-method of variable μ-propagation vector ; Velocity estimation using : d) initial S-method and e) CS based S-method

If we denote the desired signal as **x**, measurement vector as **y**, the measurement and transform matrices as **Φ** and **Ψ** respectively, then the relation follows:

$$\mathbf{y} = \mathbf{\Phi x} = \mathbf{\Phi \Psi X} = \mathbf{A X}, \tag{66}$$

where **X** corresponds to the STFT coefficients at certain available time instant $t_{si}$. To find **x** or its spectral representation **X** from an incomplete measurement vector **y**, the following optimization problem can be used:

$$\min \|\widehat{\mathbf{X}}\|_1 \quad \text{subject to} \quad \mathbf{y} = \mathbf{A}\widehat{\mathbf{X}}, \tag{67}$$

performed for each available time instant.

The results obtained by using real video sequence are shown in Fig. 10. The percentage of the available frames is 40%, due to the compressive acquisition. The moving of metronome's pendulum is observed and some of the frames from the video sequence are shown in Fig. 10a. The S-method of the µ-propagation vector calculated using the available samples, is shown in **Fig. 10**b, while the CS-based S-method is shown in **Fig. 10**c. The corresponding velocity estimations graphs are shown in **Fig. 10**e and f. It is shown that the initial S-method produces error in velocity estimation, while precise results are obtained by using the CS-based method.

*4.6.    CS in watermarking*

(A)    Data protection in terms of CS has been discussed in [136]-[141]. Fakhr in [137] proposed a watermark embedding and recovery technique based on the CS framework, tested under MP3 compression. The sparsity of both, the host and the watermark signal is assumed. The watermark is embedded into the measurement vector **y**. If we denote signal with **x**, transform domain matrix as $\Im$, a sparse signal **b** as watermark of length *L*, then the random watermark creation is described as:

$$\mathbf{w} = \mathbf{\Omega b}, \tag{68}$$

where $\mathbf{\Omega}_{M \times L}$ is the random Gaussian matrix, and *M* is the measurement vector length. Matrix **Ω** is used for random expansion of the sparse vector **b**. The embedding is done as follows:

$$\mathbf{y} = \Im \mathbf{x} + a\mathbf{\Omega b}, \tag{69}$$

resulting in watermarked measurement vector. Embedding strength **a** is adapted for each frame of the audio signal as: $\mathbf{a} = 0.04\sqrt{\sum_{i=1}^{M} \mathbf{X}i^2}$, $\mathbf{X} = \Im \mathbf{x}$. The advantage of the proposed method is that, in order to recover the clean signal, the optimization problem has to be solved and thus, matrix **Ω** has to be known. In this paper, for the optimization problem solving three methods are used: *Direct Justice Pursuit*, *Multiplying by the Inverse of* Ω and *Multiplying by the annihilator of* Ω.

(B)    An image watermarking procedure in the CS scenario is proposed in [139]. The randomly chosen pixels that serve as CS measurements are used to bring the watermark. The image is firstly divided into the blocks and measurements are selected from each block. Samples are taken from the space domain, while the image sparsity is assumed in the DFT domain.

If we denote the *N*×*N* image block as $\mathbf{I}_j$, vector of measurements for *j*-th block as $\mathbf{y}_j$, $\mathbf{T}_j$ vector of transform domain coefficients (DFT) of the block $\mathbf{I}_j$, **Ψ** as the Fourier transform matrix and $\mathbf{\Omega}_j$ as the measurement matrix for the block *j*, then the measurement vector is defined as:

$$\mathbf{y}_j = \mathbf{\Omega}_j \mathbf{I}_j = \mathbf{\Omega}_j \mathbf{\Psi} \mathbf{T}_j. \tag{70}$$

The watermarked measurement vector $\hat{\mathbf{y}}_j$ is obtained as follows:

$$\hat{\mathbf{y}}_j = \mathbf{y}_j + \mu \omega_j, \tag{71}$$

where *μ* denotes watermark strength and *ω* is *M*×1 watermark vector (*M* denotes the number of measurements). The vector of watermarked coefficients is used to recover the image according to the total variation optimization:

$$\min_{\mathbf{T}} TV(\mathbf{T}_j) \text{ subject to } \mathbf{y}_j = \mathbf{\Omega}_j \mathbf{\Psi} \mathbf{T}_j . \tag{72}$$

The reconstructed image block $\mathbf{I}_{Rj}$ is obtained as $\mathbf{I}_{Rj} = \mathbf{\Psi}\mathbf{T}_j$. Watermark detection is based on using the standard correlator that requires measurement matrix $\mathbf{\Omega}_j$ to be known:

$$D(\boldsymbol{\omega}) = \sum_i y_{i\omega} \omega_i . \tag{73}$$

The procedure is tested on 256×256 image, divided into 16×16 block. From each block, 50% of the pixels is randomly chosen and serve as a measurement in the reconstruction process, and carry watermark as well. The results are shown in Fig. 11. PSNR between original and watermarked/reconstructed image is 31.79 dB.

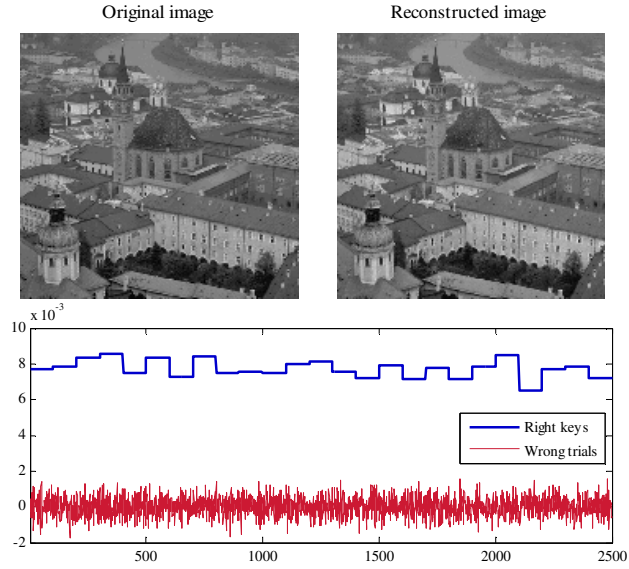

Fig. 11 : a) Original image; b) Watermarked and reconstructed image; c) Detector responses for 25 right keys and 2500 wrong trials (100 wrong trials for each right key)

## 5. CONCLUSION

The paper focuses on the Compressive Sensing, as an approach that records an intensive development in signal processing in recent years. An overview of the Compressive Sensing applications and commonly used algorithms for reconstruction of the signals with missing data, is given. Algorithms for the reconstruction of both, 1D and 2D signals, are described in the paper. The paper covers the applications starting from the radar signal processing, communications, biomedical signals and image reconstruction, through natural image reconstruction, velocity estimation in video signal processing, CS-based protection of the digital data and hardware devices designed based on the CS principles. Experimental results are provided in order to show the performance of the presented algorithms and approaches.

**Acknowledgement**: *The paper is a part of the research supported by the Montenegrin Ministry of Science, project grant: "New ICT Compressive sensing based trends applied to: multimedia, biomedicine and communications (CS-ICT)" (Montenegro Ministry of Science, Grant No. 01-1002). The authors are thankful to Dr Josip Musić for testing the performance of the object detection in Search&Rescue images.*